\documentstyle[amssymb,preprint,aps]{revtex}

\begin{document}
\title{Cellular Structures for Computation in the Quantum Regime}
\author{S. C. Benjamin and N. F. Johnson}
\address{Centre for Quantum Computation, Clarendon Laboratory, University of Oxford,\\
UK.}
\date{\today}
\maketitle

\begin{abstract}
We present a new cellular data processing scheme, a hybrid of existing
cellular automata (CA) and gate array architectures, which is optimized for
realization at the quantum scale. For conventional computing, the CA-like
external clocking avoids the time-scale problems associated with
ground-state relaxation schemes. For quantum computing, the architecture
constitutes a novel paradigm whereby the algorithm is embedded in spatial,
as opposed to temporal, structure. The architecture can be exploited to
produce highly efficient algorithms: for example, a list of length $N$ can
be searched in time of order $\sqrt[3]{N}$.
\end{abstract}

\newpage

There has been much recent interest in the topic of information processing 
\cite{Feynman} at the nanometer scale where quantum mechanical effects can
play an important role \cite{NewScientist}. Theoretical design schemes have
been reported for the regimes of conventional, classical computation\cite
{domino,lines,Obermayer,ACApaper} and, more recently, quantum computation
utilizing wavefunction coherence across the entire structure \cite{Steane}.
Many of these designs have in common the feature that they are formed from
many simple units which interact only locally. In this respect they are
reminiscent of the mathematical objects called cellular automata (CA's),
regular arrays of locally interacting cellular units driven by global update
rules \cite{ToffMargolus}. It is known that when such structures are defined
in suitable terms, they can in principle perform both conventional \cite
{ToffMargolus}\ and quantum computing \cite{SethCA}. However, only
for the simple case of a one-dimensional CA \cite{SethCA} is it understood
how best to implement computation in a real physical system formed of directly
interacting, globally driven cellular units.

Here we exploit existing ideas of cellular automata and conventional
gate-arrays to form a new cellular scheme which is optimized to function
under real physical conditions. Such conditions may include long-range
cell-cell interactions (eg. Coulombic) , cells possessing only two stable
states, and the inability of a cell to distinguish its neighbors. In
operation the scheme would offer significant advantages over other
nanocomputing proposals. For nanometer-scale conventional computing, the use
of externally applied updates implies that the device is driven through a
definite set of internal states on a well-defined time-scale. By contrast,
the most popular comparable scheme \cite{domino,scienceQCA} relies on
thermal relaxation to the ground-state of the system; hence at any non-zero
temperature there is the danger that the system's evolution may reverse or
become stuck in a computationally meaningless metastable state \cite
{proceedings,mitre}. Furthermore we could choose cells whose internal states
are well separated in energy, thereby making room temperature operation
feasible. For quantum computing, our scheme offers advantages over
implementations such as the ion trap \cite{Steane} or the simple
one-dimensional CA \cite{SethCA} because it allows many gate operations to
be performed simultaneously (and at points irregularly distributed over the
structure). Also for both conventional and quantum computation our
architecture can process a {\em series} of many independent inputs
simultaneously; this is `pipe-lining' taken to its ultimate limit. These
advantages can be exploited to produce, for example, an enhanced quantum
searching algorithm \cite{Grovers}.

We first define the network architecture. This architecture has a number of
possible physical realizations \cite{longPaper}; later we give two examples.
The network consists of many individual simple units called `cells'. Each
cell has two distinct internal states, say `0' and `1', in the energy range
of interest. The state of a cell can be changed, conditional on the cell's
current state and the states of its neighboring cells. It is not necessary
(or desirable) to address one cell at a time; instead the entire structure
is subject to a conditional update `rule' during which those cells that meet
the condition will change their state. The cells in the network are only
required to be sensitive to the states of their {\em nearest} neighbors: we
will take the term `neighbors' to mean `nearest neighbors'. Neighboring
cells will never simultaneously meet an update condition. Cells are not
required to be able to distinguish one neighbor from another. This is a very
advantageous feature because many fabrication techniques would naturally
produce equally spaced units, and moreover the requirement of
distinguishability would severely restrict the suitable forms of physical
interaction (see later). We employ a number of different `types' of cell,
where `type' denotes a subset of cells which have the same energy separation
between `0' and `1'. With a greater number of types, functions can be
realized by more compact networks. Conversely, using more elaborate networks
allows certain functions to be performed with only a single cell type \cite
{longPaper}. Here we present networks that represent a good trade-off
between the number of cell types and the network's complexity. A Java Applet
is available \cite{applet} for verifying the properties of our networks and
for designing new networks.

We first implement the elementary functions {\em transportation of data}, 
{\em fanning-out (i.e. copying) of data}, and the {\em logical operations}
XOR and NOR; this complete set of components \cite{Feynman} suffices to
produce the cellular equivalent of any conventional computational circuit.
Figure 1 shows how just two types of cell, $\alpha $ and $\beta $, can be
arranged to produce these functions. Data bits, labeled by $x_{1}$ etc.,
move through the networks in response to a certain repeating sequence of
conditional updates. For each update we employ the notation $\stackrel{%
t\rightarrow u}{{w_{v}}}$ to denote the following: cells of type ${w}$ which
are presently in state $t$ will change to state $u$ if and only if the
`field' is of strength $v$; the `field' is defined as the number of nearest
neighbors in state `1' minus the number in state `0'. Hence $\stackrel{%
1\rightarrow 0}{{\beta _{-2}}}$ indicates that $\beta $ cells whose current
state is `1' are to change their state to `0' if, and only if, two more
neighbors are in state `0' than are in state `1'. The `field' is the proper
control variable since the cells will be `aware' of their neighbors only
through the net effect of, for example, their electrostatic fields. The
following master sequence of updates suffices to drive data through any and
all of the networks in Fig. 1: $\stackrel{0\rightarrow 1}{{\beta _{0}}}$,$%
\stackrel{1\rightarrow 0}{{\alpha _{0}}}$,($\stackrel{0\rightarrow 1}{{%
\alpha _{-1}}}$,$\stackrel{0\rightarrow 1}{{\beta _{-3}}}$),($\stackrel{%
1\rightarrow 0}{{\beta _{0}}}$,$\stackrel{1\rightarrow 0}{{\beta _{-2}}}$),$%
\stackrel{0\rightarrow 1}{{\alpha _{0}}}$,($\stackrel{1\rightarrow 0}{{%
\alpha _{-1}}}$,$\stackrel{1\rightarrow 0}{{\alpha _{1}}}$,$\stackrel{%
1\rightarrow 0}{{\beta _{-1}}}$). Here brackets indicate a sub-sequence of
updates which can be performed in any order, or simultaneously. Since this
sequence will drive {\em any} network formed from the components of Fig. 1,
it is straightforward to implement any function for which a conventional
gate array is known. Figure 2 provides an example: binary addition. The
structure shown in Fig. 2(d) is geometrically a regular lattice; the
algorithm is embedded in the choice of cell types rather than the position
of the cells. One could implement any algorithm by taking a region of a
perfectly regular hexagonal lattice and selectively assigning a new `type'
to particular cells. This suggests an efficient means of manufacture. If the
cell types could subsequently be re-assigned (eg. if the type were
determined by a local electrostatic gate) then the structure could be
programmed for different algorithms. Furthermore, such programmability would
allow a newly fabricated device to be configured to test the integrity of
the cells - defective cells could then be routed around. Defect tolerance
may be a fundamental requirement for a successful nanocomputing scheme \cite
{HPterra}.

The networks shown in Fig. 1 each contain three independent sets of bits
(e.g. \{$z_{1}$\}, \{$x_{2}$,$y_{2}$\}, \{$x_{3}$,$y_{3}$\} in Fig. 1(d)); a
network formed from these components with a total `depth' of $N$ cells can
simultaneously process $\frac{1}{3}N$ sets of bits. This property, a kind of
ultra-dense `pipe-lining', is clearly very advantageous for certain problems
such as numerical integration in which the same function must be applied to
many inputs. In general if an algorithm which is $n$ gates `deep' must be
repeated $m$ times, then a simple computer will take time of order $nm$, but
the cellular computer will require only of order $n+m$ repetitions of the
update sequence (order $n$ repetitions for the first answer to appear, then
another appears with {\em each} subsequent repetition). This is explored
later.

To implement a conditional update, we rely on a cell's transition energy $%
\omega $ being perturbed by the state of its neighbors: $n$ distinguishable
neighbors would divide $\omega $ into $2^{n}$ levels, each of which would be
further split by the effect of the many non-neighboring cells. In order to
drive a transition in reasonable time we should address these sub-levels
collectively, ie. in bands. We will obtain finite bands for any $d$%
-dimensional array if the range of the interaction energy $g(r)$ is shorter
than $r^{-d}$. However there is a requirement which is more severe and more
complex: {\em bands must not overlap}. If neighbors need not be
distinguishable (as in the schemes here), then certain bands {\em are}
allowed to overlap. For the design shown in Figs. 1 \& 2 the form $%
g(r)\varpropto r^{-3}$ is sufficiently short \cite{rm3But}. The quantum dot
cells mentioned below have an interaction of this form. In order to assign a 
$g(r)$ to previous CA computing schemes \cite{ToffMargolus} they must first
be made suitable for direct physical realization; we find that such
modifications simply yield less evolved versions of the ideas presented here.

So far we have discussed only conventional, irreversible computation
employing two-input, one-output gates. Our updates were generally
irreversible, and hence non-unitary, because they addressed cells of a given
state. To see this, consider a simple line of cells ...$\alpha $$\beta $$%
\alpha $$\beta $$\alpha $... in which one $\alpha $ cell is in state `1' and
all the other cells are in state `0'. The update $\stackrel{1\rightarrow 0}{%
\alpha {_{-2}}}$ will change the `1' to a `0' {\em without altering any of
the other cells}, i.e. it will erase the information represented by the
position of the `1'. We now introduce a new update: the notation $w_{v}^{{U}%
} $ means that each cell of type $w$ is subjected to the unitary transform ${%
U} $ if and only if the `field' (defined above) is of strength $v$; when we
omit ${U}$ an inversion (ie. a NOT) is implied. We can immediately reverse
such an update by applying $w_{v}^{{U}^{\dagger }}$\ (recall that a cell and
its neighbors are never changed by the same update). Figure 3(a) shows a
line of cells that can act as a wire when subject to these updates. With
each single bit of data represented by the states of a pair of adjacent
cells (i.e. 00 or 11), the short sequence $\beta _{0}$,$\alpha _{0}$ is
sufficient to move all the bits along the wire by two cells. This change
could {\em not} have been accomplished by the class of updates used earlier.

We could use these unitary updates to describe a classical reversible
cellular computer \cite{longPaper}, but instead we proceed to the general
case of a cellular quantum computer (QC). Quantum computation is a
relatively new paradigm which holds the promise of fundamentally superior
performance \cite{Steane,deutsch}. The implementation of the wire shown in
Fig. 3(a)\ remains valid when we generalize the bits $x_{i}$ to `qubits' $%
x_{i}=A|0\rangle +B|1\rangle $, which would be represented by a pair of
cells as $A|00\rangle +B|11\rangle $. One set of gates from which a general
QC could be built consists of a number of one-qubit transformations, which
can be thought of as quantum generalizations of the NOT\ gate, and a
particular two-qubit gate called the control-not (CNOT) \cite
{elementaryGates}. In Figs. 3(b) and (c) we show possible implementations
for these gates. The following master sequence \cite{betterthan21}, $\beta
_{0}$,$\alpha _{-1}$,$\alpha _{0}$,$\alpha _{1}$,$\beta _{-1}$,$\alpha _{0}$%
,($\alpha _{3}$,$\alpha _{1}$),$\alpha _{0}$,($\gamma _{0}$,$..$),$\beta
_{0} $,($\beta _{-1}$,$\gamma _{-2}^{{U}}$, $..$),$\alpha _{0}$,($\gamma _{0}
$,$.. $),$\beta _{0}$,($\alpha _{1}$, $\alpha _{3}$),$\beta _{0}$,$\beta
_{-1}$,$\alpha _{1}$,$\beta _{0}$,$\alpha _{-1}$,$\alpha _{0}$, will operate
all the components shown in Fig. 3 (along with additional one-qubit gates if
their corresponding updates are inserted as indicated by `..'). Therefore
this sequence will drive a general QC formed from those components. Our
earlier remarks, concerning the possibility of growing a regular array and
subsequently programming it by setting cell types, apply equally well here.
However since QC's will probably be built to attack very specific problems,
programmability may be a less important feature. The network shown in Fig.
3(c) is more dense that those of Fig. 1 and consequently the cell-cell
interaction must be shorter range; simulation shows that $r^{-4}$ suffices.

Previous QC proposals \cite{Steane,SethCA} typically employ a simple
periodic spatial topology (e.g. a 1-dimensional array of units) together
with a complex irregular temporal sequence of operations. By contrast, the
present scheme involves a complex, irregular spatial topology together with
a simple, periodic temporal sequence. This alternative paradigm can offer
advantages both in speed, due to many gate operations being performed
simultaneously, {\em and} in the potential for dense `pipe-lining'. We will
now demonstrate the consequence of these features using the specific case of
a quantum searching algorithm (QSA) \cite{Grovers}. Suppose that we have a
list of $N$ values and it is known that just one of them, $x$, satisfies
some condition $f(x)=1$. If we try to find this unique value by searching
the list on a classical computer, the expected time to complete the task
will be of order $k_{1}N$, where $k_{1}$ is the time to evaluate the
function $f$ once \cite{Grovers}. If instead we have a QC such as an ion
trap, we could use a QSA\ to find $x$ in an expected time of order $k_{2}%
\sqrt{N}$, where $k_{2}$ is the time to perform one `Grover iteration' (each
Grover iteration calls for the function $f$ to be evaluated once) \cite
{Grovers}. Applying $r$ Grover iterations to an initial state $\frac{1}{%
\sqrt{N}}\Sigma _{i=1}^{N}|i\rangle $ causes the amplitude of $|x\rangle $ to
become $\sin ((2r+1)\theta )$, with $\theta $ defined by $\sin \theta =\frac{%
1}{\sqrt{N}}$, $0<\theta <\frac{\pi }{2}$\cite{Grovers}. On a
one-dimensional QC such as an ion trap, one can do little better than simply
performing $r\simeq \frac{\pi }{4\theta }$ ($\approx \frac{\pi }{4}\sqrt{N}$
for large $N$) iterations to maximize the amplitude, and then making a
single measurement \cite{Grovers}. However, using our network QC with the
strategy described below will reduce the expected time to below $\sqrt[3]{%
k_{3}^{2}k_{1}N}$, where $k_{3}$ is the time to perform one Grover iteration
on our cellular network. Typically we may expect $k_{3}$ as a function of$\
N $ to be no worse than $k_{2}$ \cite{knoWorse}; for many functions $f$ of
interest $k_{1}$,$k_{2}$ and $k_{3}$ would all be merely logarithmic in $N$.
In the following we will assume that $N$ is large, so that 
$\theta \approx\frac{1}{\sqrt{N}}$.\ The cellular network would be configured so that
it implements only $r$ Grover iterations, with $r\ll \sqrt{N}$. Then when we measure
the output, we will obtain the desired $x$ with only a small probability $\varepsilon
\approx \frac{(2r+1)^{2}}{N}$. However, because of the pipe-lining effect
explained above, after the first measurement we can make another independent
measurement after {\em each} repetition of the update sequence. If one
repeatedly makes independent attempts at some task, with each attempt having
a probability\ of success $\varepsilon $, the expected number of attempts
required to succeed is just $\frac{1}{\varepsilon }$. Thus the expected time
to find $x$ is $k_{3}r+\frac{N}{(2r+1)^{2}}k_{1}$ (the $k_{1}$ factor
accounts for the fact that we must apply $f$ to each measured value to see
if it is the desired $x$). This expression is minimized by $2r+1=\sqrt[3]{%
4k_{3}^{-1}k_{1}N}$ (which for reasonable $k_{1}/k_{3}$ satisfies our
original assumption that $r\ll \sqrt{N}$), the minimum being $\frac{3}{4}%
\sqrt[3]{4k_{3}^{2}k_{1}N}-\frac{1}{2}k_{3}$. It is also interesting to note
that if we are satisfied with only {\em matching} the speed of the simple
QC, then our network need implement only $r\sim \sqrt[4]{N}$ Grover
iterations. This is important since it means that coherence need only be
maintained over a vastly smaller number of steps (recall that there is no
quantum entanglement `along' the pipe line, ie. in Fig. 3 the sets of
variables with different subscripts are independent). Finally we note that
for the related problem of a list containing $t$ solutions, {\em all} of
which we wish to find, the performance comparison can be even more dramatic
(taking $t$ of order $\sqrt[3]{N}$, if the simple QC requires time $k_{2}S$
then our network QC\ requires only time $2k_{3}\sqrt{S}$, neglecting
logarithmic factors \cite{longPaper}).

We now mention two potential physical realizations of the cell. The first is
a bistable double quantum-dot driven through its internal states by laser
pulses. This structure is fully described in Ref. \cite{Obermayer}, here we
will simply remark on its features. The device is bistable and can be
switched either by pumping (which would implement our irreversible updates
of \ form $\stackrel{1\rightarrow 0}{\alpha {_{0}}}$ ) or by coherent
stimulation (which would implement the reversible updates $\alpha {_{0}}$\
). The switching energy is near the ideal minimum for a device that is
stable at room temperature. The cell-cell interaction is of the form $r^{-3}$%
. Rapid decoherence may restrict this implementation to non-QC applications,
although photonic band gap engineering might alleviate this. The second
implementation derives from the solid-state QC scheme recently proposed by
Kane \cite{Kane}. In this scheme, the cells are the spin-$\frac{1}{2}$
nuclei of $^{31}$P impurity atoms embedded in Si and subject to an external
magnetic field. The exponential form of the effective interaction between
the nuclei is shorter than $r^{-4}$ and so is suitable for the QC scheme
presented here. The interaction is not diagonal in the computational basis 
\cite{Kane}, however when we introduce the principle of adjacent cells being
of different types, the effect of the off-diagonal terms disappears \cite
{longPaper} . A cell's type would be determined by electrostatic gates just
as in the original proposal, however the second set of gates (`J gates') are
not needed.

In conclusion we have presented a new architecture hybridised from existing
CA and gate-array architectures and optimized for nanometer scale
realization. The architecture offers clear advantages for both conventional
computing and true quantum computing.

The authors wish to thank W. van Dam, M. Mosca, A. Ekert and G. Mahler for
useful discussions. This work was funded by an EPSRC Grant for Photonic
Materials.

\newpage

\noindent \noindent \centerline{\bf Figure Captions}

\bigskip

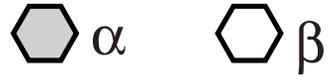
\noindent {\bf Figure 1}: Cellular network components. Cells drawn with
dashed borders will remain in state `0' at all times. Data bits are
represented by $x_{1}$, etc.(a) A wire before (i) and after (ii) the
application of an update sequence. When the update $\stackrel{0\rightarrow 1%
}{{\beta _{0}}}$ is applied to (i) the data bits will be copied onto the $%
\beta $\ cells, and the update $\stackrel{1\rightarrow 0}{{\alpha _{0}}}$
will erase the original bits. Similarly the updates $\stackrel{0\rightarrow 1%
}{{\alpha _{-1}}}$,$\stackrel{1\rightarrow 0}{{\beta _{0}}}$ will move the
bits one cell further down, and $\stackrel{0\rightarrow 1}{{\alpha _{0}}}$, $%
\stackrel{1\rightarrow 0}{{\alpha _{-1}}}$ will complete the cycle to yield
(ii). Each repetition of the sequence $\stackrel{0\rightarrow 1}{{\beta _{0}}}
$,$\stackrel{1\rightarrow 0}{{\alpha _{0}}}$,$\stackrel{0\rightarrow 1}{{%
\alpha _{-1}}}$,$\stackrel{1\rightarrow 0}{{\beta _{0}}}$,$\stackrel{%
0\rightarrow 1}{{\alpha _{0}}}$,$\stackrel{1\rightarrow 0}{{\alpha _{-1}}}$
will move the bits three cells further. (b) A network which performs the 
{\em copy} or {\em fanout} operation: one stream of data is divided into two
identical copies. (c) \& (d) Networks to perform the logical operations NOT
and NOR respectively; the network for XOR operation is identical to NOR
except that the central cell is of type $\alpha $. The master sequence,
given in the text, drives any and all of these networks.

\noindent

\noindent {\bf Figure 2}: (a) A gate array for the elementary 1-bit adder
(or `half-adder') formed from NOT, XOR\ and NOR gates. (b) Implementation
using the components of Fig.1. (c) One way of producing an 3-bit full adder
using 1-bit adders and XOR's as components. (d) The corresponding cellular
version. Pipe-lining enables this network to simultaneously process 15
additions.

\noindent

\noindent \noindent {\bf Figure 3}: (a) A `wire' driven by a sequence of
just two updates: $\beta _{0}$,$\alpha _{0}$. (b) One way of effecting a
given one-qubit transformation ${\bf U}$ by introducing a further cell-type $%
\gamma $: if the two cells labeled $x_{2}$ are in the state $A|00\rangle
+B|11\rangle $ then the sequence $\beta _{0}$,$\alpha _{0}$,$\gamma _{0}$,$%
\beta _{0}$, $\gamma _{-2}^{{U}}$, $\alpha _{0}$,$\gamma _{0}$,$\beta _{0}$,$%
\alpha _{0}$, will move and transform the qubit so that it is represented by
a $\beta $,$\alpha $ pair in the state $C|00\rangle +D|11\rangle $ with $%
{\textstyle{C \choose D}}%
$ related to $%
{\textstyle{A \choose B}}%
$ by ${\bf U}$. Other one-bit transformations can be implemented by
employing additional cell types, $\gamma $,$\delta $,$\epsilon $.. \cite
{2forQC}. (c) The truth table and cellular network for the two-bit CNOT
gate. Unmarked cells are in state `0'. One possible \cite{betterthan21}
update sequence to drive it is: $\beta _{0}$,$\alpha _{-1}$,$\alpha _{0}$,$%
\alpha _{1}$,$\beta _{-1}$,$\alpha _{0}$,( $\alpha _{3}$, $\alpha _{1}$),$%
\alpha _{0}$, $\beta _{0}$,$\beta _{-1}$,$\alpha _{0}$,$\beta _{0}$,($\alpha
_{1}$ , $\alpha _{3}$),$\beta _{0}$,$\beta _{-1}$,$\alpha _{1}$,$\beta _{0}$,%
$\alpha _{-1}$,$\alpha _{0}$.


\begin{references}
\bibitem{rm3But}  In order to accept $r^{-3}$ we have to reduce density of
data along the `pipe-line' by a half, ie. the sets of bits are separated by
6 cells.
\bibitem{betterthan21}  There may exist shorter sequences that perform the
same function; the quoted sequence is the shortest that the authors have
found so far.
\bibitem{knoWorse}  It seems that the ability of the cellular network to
perform gate operations in parallel more than compensates for an ion trap's
ability to operate on non-adjacent qubits. This easily shown for the
transformation $U_{0}$\ which is key to the Grover iteration \cite{longPaper}.
\end{references}
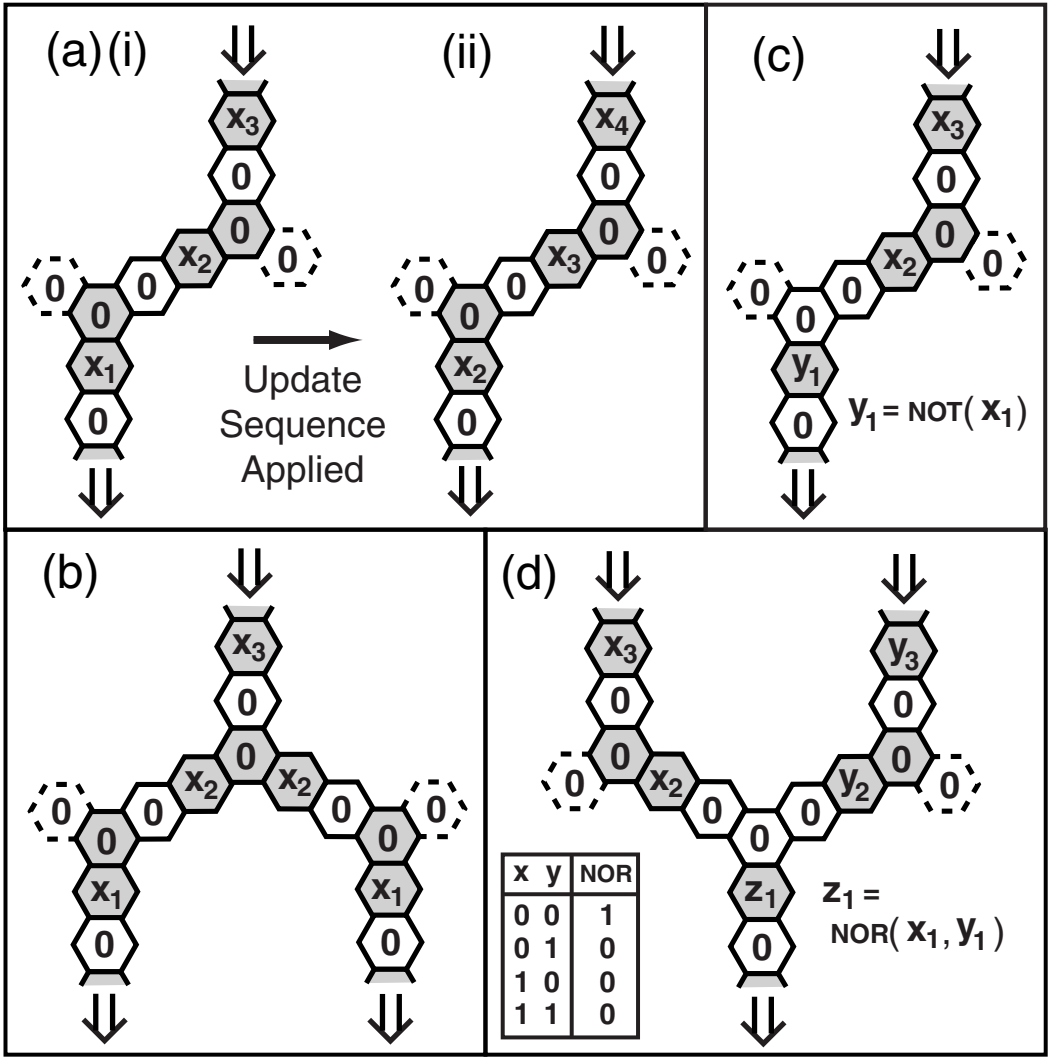
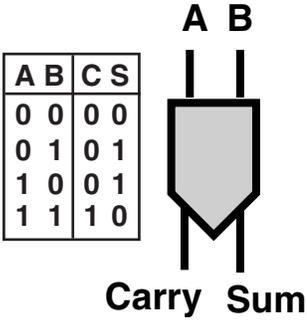
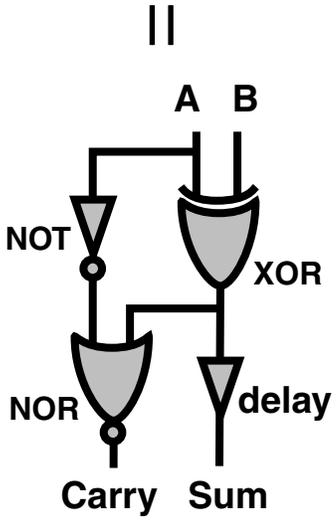
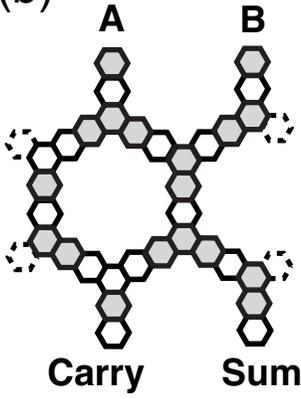
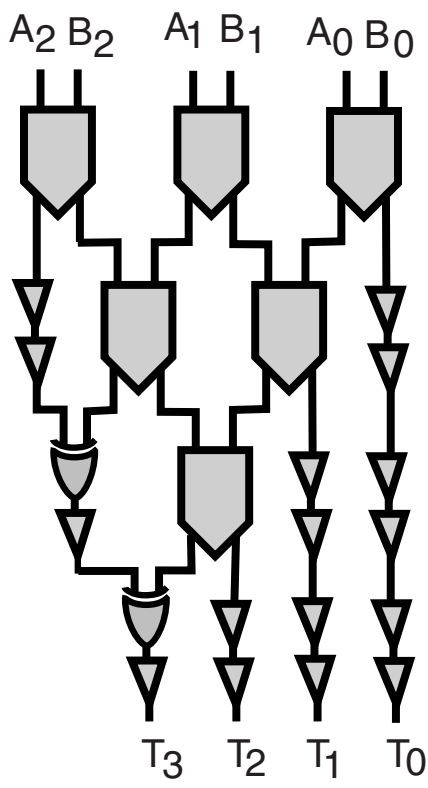
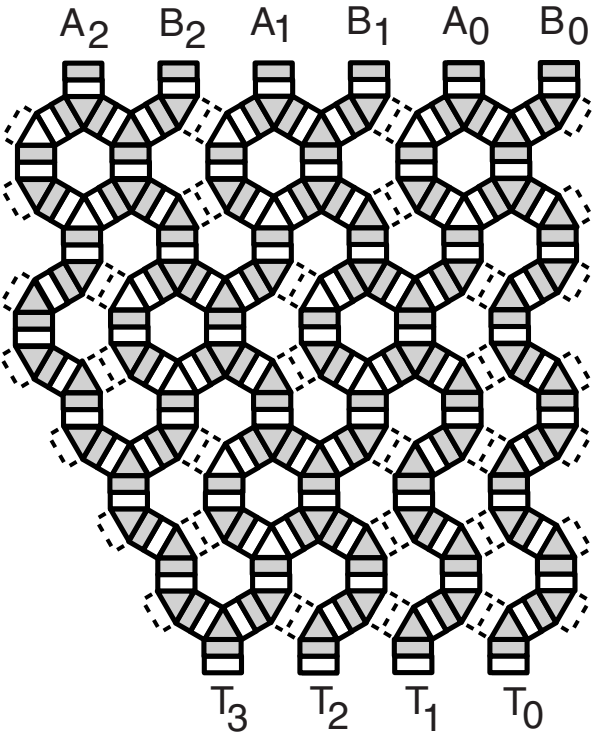
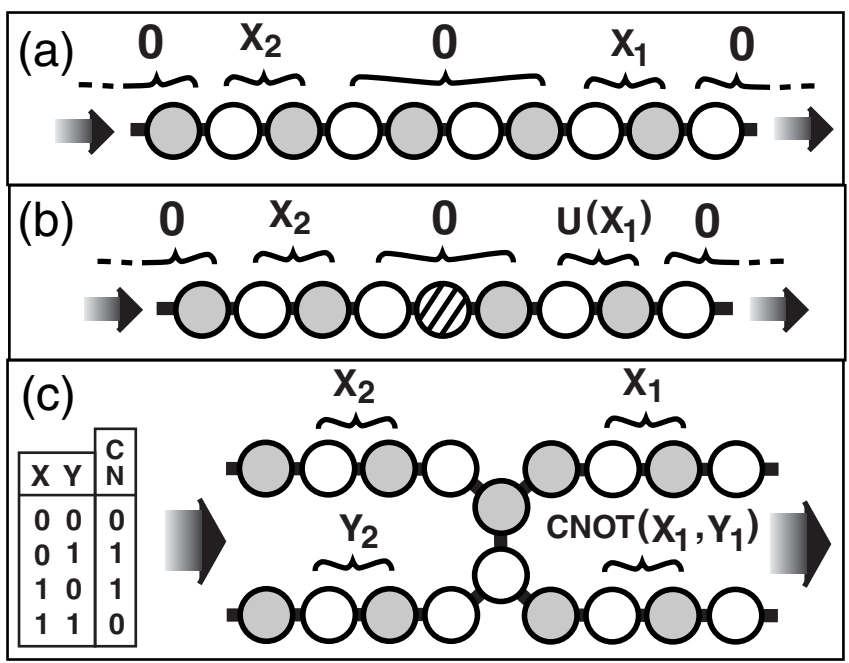
\end{document}